\documentstyle[twoside,epsfig,fleqn,espcrc2]{article}

\def\Pom{{I\!\!P}}
\def\Reg{{I\!\!R}}

\def\Por{{\tilde{I\!\!P}}}

\title{
\hskip+13cm
{\normalsize\rm SI--97--17}\\
\hskip+13cm
{\normalsize\rm November 1997}
\hskip-13cm
\vskip+1cm
HARD DIFFRACTION AND CENTRAL DIFFRACTION IN HADRON--HADRON AND
PHOTON--HADRON COLLISIONS\thanks{Talk presented by J.~Ranft at 
the International Symposium on Near Beam Physics, Fermilab, 22-24 Sept.
1997}
}
\author{R.~Engel\thanks{e--mail: eng@lepton.bartol.udel.edu}, 
University of Delaware, Bartol Research Institute, Newark, DE 19716 USA \\
 J.~Ranft\thanks{e--mail: Johannes.Ranft@cern.ch},
FIGS and Physics Dept. Universit\"at Siegen, D--57068 Siegen, Germany}

\begin{document}

\begin{abstract}  
Hadron production in single and central diffraction
dissociation is studied in a model which includes soft hadron interaction
as controlled by a supercritical pomeron parametrization
and hard diffraction. Hard
diffraction is described using leading-order QCD matrix elements together
with the parton distributions for the proton, the less well known photon
parton densities and a conjectured parton distribution function for the
pomeron.  Within this model, particle production in collisions with
pomerons exhibit properties like multiple soft interactions and multiple
minijets, quite similar to hadron production in non-diffractive hadronic
collisions at high energies.  However, important differences occur in
transverse momentum jet and hadron distributions.  It is shown that the
model is able to describe data on single diffractive hadron production
from the CERN-SPS collider and from the HERA lepton-proton collider as
well as first data on central diffraction dissociation. We
present also model predictions for single and central
diffraction at TEVATRON.
\end{abstract}
 \maketitle

\section{Introduction}

High-energy hadron production
in hadron--hadron collisions and in
hadronic interactions of photons is characterized by two mechanisms:
(i) minijet production and (ii)
soft hadronic interactions. Whereas the minijet cross section
can be estimated applying the QCD-improved parton model, soft
hadron production cannot be computed
directly from perturbative QCD.
Most models for multiparticle production being
constructed in form of Monte Carlo
event generators use soft and hard mechanisms. Such models are
usually called minijet models if they use minijets and
a simple model
for the soft component of the interaction. They are called two
component Dual Parton models (DPM's) if they use minijets and
incorporate  a evolved soft component
which is derived from Regge theory, Gribov's reggeon
calculus \cite{Gribov67a-e,Gribov68c-e}
and Abramowski-Gribov-Kancheli (AGK) cutting
rules \cite{Abramovski73-e} (a review is given in Ref.\cite{Capella94a}).

Models inspired by Regge theory or the DPM describe
high-mass diffractive hadron production in terms of the so-called
triple-pomeron graph.
According to this diffractive processes can be considered as
collisions of a color neutral object, the pomeron, with hadrons,
photons or other pomerons.
Experimental data on diffraction support this idea showing that diffraction
dissociation exhibits similar features as non-diffractive hadron
production whereas the mass of the diffractively produced system
corresponds to the collision energy in non-diffractive interactions
\cite{Goulianos83,Bernard86a}.
The striking similarities between diffractive and
non-diffractive multiparticle production suggest that multiple soft and
hard interactions may also play an important role in high-mass
diffraction dissociation.

However the pomeron cannot be considered as
an ordinary hadron. It is important to keep in mind that the pomeron
is only a theoretical object 
providing an effective description of the important 
degrees of freedom of a certain sum of Feynman diagrams in Regge limit. 
Pomeron-hadron
or pomeron-pomeron interactions can only be discussed in the framework of
collisions of other particles like hadrons or photons in terms of
single, double or central diffraction dissociation.

The DPM was already successfully applied to diffractive hadron
production reactions \cite{Innocente86,Ranft87c,Roesler93} 
and even hard diffractive
processes \cite{Engel95c}. In \cite{Engel96e} cross sections on single
and central diffraction were calculated. Up to now,
the minijet component in diffractive
processes within the two-component DPM 
was obtained using a parton distribution function (PDF)
for the pomeron and flux factorization. 
The soft component of diffractive interactions
was described by two hadronic chains (cutting the triple-pomeron graph).
Here we will argue, that for the description of 
diffraction dissociation producing hadronic systems with very large
masses, such
models are not enough. Also for high-mass diffractive hadron production
we need multiple soft and multiple hard interactions.


\section{The Model\label{the-model}}

\subsection{The event generator {\sc Phojet}}

In the {\sc Phojet} model\cite{Engel95a,Engel95d}, 
interactions of hadrons are described
within the DPM in terms of
reggeon ($\Reg$) and pomeron ($\Pom$) exchanges. 
The realization of the DPM with a hard and a soft
component is similar to the event generator {\sc Dtujet}
\cite{Aurenche92a,Bopp94a} for $p$--$p$ and $\bar p$--$p$ collisions. 
In the
following we briefly describe the  treatment of the pomeron exchange
in non-diffractive interactions
since the same framework is also used for the description of particle
production in diffraction dissociation.

The pomeron exchange is artificially
subdivided into  {\it soft}
processes and processes with at least one large momentum
transfer ({\it hard} processes).
This allows us to use the predictive power of the QCD-improved Parton
Model with
lowest-order QCD matrix elements \cite{Combridge77,Duke82a} and
parton density functions.
Practically, soft and hard processes are distinguished by applying
a transverse momentum cutoff $p_\perp^{\mbox{\scriptsize cutoff}}$ to
the partons. Consequently, the pomeron is considered as a 
two-component object with
the Born graph cross section for pomeron exchange given by 
the sum of hard and soft cross sections.


\subsection{Diffractive cross section calculation}


Concerning diffraction dissociation, our approach is the following.

In order to get an effective parametrization of Born graphs describing
diffraction within Gribov's reggeon calculus, we calculate the triple-,
loop- and double-pomeron graphs using a renormalized pomeron intercept 
$\alpha_\Por = 1+\Delta_\Por = 1.08$.
For example, let's consider the the Born graph cross sections for 
high-mass diffraction dissociation in $A$--$B$
scattering (for simplicity, we omit in the following expressions the
pomeron signature factors; for a discussion of the couplings etc.\ see
\cite{Engel96e}).

High-mass single diffraction dissociation of particle $A$
is calculated using the triple-pomeron approximation 
\begin{eqnarray}
\frac{d^2\sigma^{\rm TP,A}_{AB}}{dt\,dM_{\rm D}^2} &=&
\frac{1}{16 \pi} 
\left(g^0_{B\Pom}\right)^2\ g^0_{3\Pom}
\ g^0_{A \Pom}
\left(\frac{s}{s_0}\right)^{2\Delta_\Por}
\nonumber\\
& &\times
\left(\frac{s_0}{M_{\rm D}^2}\right)^{\alpha_\Por(0)}
\exp\left\{b_{AB}^{\rm SD}\ t\right\}.
\label{triple-Pom}
\end{eqnarray}
The differential cross sections for the high-mass double diffraction
dissociation  reads
\begin{eqnarray}
\frac{d^3\sigma_{\rm LP}}{dt\,dM_{D_1}^2\,dM_{D_2}^2} &=&
\frac{1}{16 \pi}
g^0_{A\Pom}\ \left(g^0_{3\Pom}\right)^2
\ g^0_{B\Pom}
\nonumber\\
& \times&
\left(\frac{s}{s_0}\right)^{2\Delta_\Por}
\left(\frac{s_0}{M_{D_1}^2}\right)^{\alpha_\Por(0)}
\nonumber\\
& \times&
\left(\frac{s_0}{M_{D_2}^2}\right)^{\alpha_\Por(0)}
\exp\left\{ b^{\rm DD}_{AB}\ t \right\}.
\label{loop-Pom}
\end{eqnarray}
Finally, we give the expression for central diffraction dissociation
\begin{eqnarray}
\frac{d\sigma_{\rm DP}}{dt_1 ds_1 dt_2 ds_2} 
&=& \frac{1}{256 \pi^2} \frac{1}{s_0} ( g_{A\Pom}^0 g_{B\Pom}^0
g_{3\Pom}^0 )^2
\nonumber\\
&\times&
\left(\frac{s}{s_0}\right)^{\Delta_\Por}
\left(\frac{s}{s_1}\right)^{\Delta_\Por}
\left(\frac{s}{s_2}\right)^{\Delta_\Por}
\nonumber\\
&\times&
\frac{1}{s_1 s_2}
\exp\left\{b^{\rm CD}_{\rm A}\ t_1 + b^{\rm CD}_{\rm B}\ t_2 \right\}.
\label{double-Pom}
\end{eqnarray}

The experimentally observable diffractive cross sections (i.e.\ cross
sections of rapidity gap events) are considerably smaller than the Born
graph cross section given in (\ref{triple-Pom}), (\ref{loop-Pom}) and
(\ref{double-Pom}). The reason for this are significant shadowing 
contributions  which are estimated by a two-channel eikonal model
\cite{Aurenche92a,Engel95d}.
It should be emphasized that these shadowing contributions lead to an
effective pomeron flux function which is energy as well as projectile
and target dependent. Hence the pomeron flux does not obey factorization within
this model.


\subsection{Particle production in diffraction dissociation}




However, not 
only for cross section calculations, but also for the description of
particle production, shadowing effects are important. Unitarity and
AGK cutting rules predict that shadowing effects are directly 
connected with so-called multiple interaction contributions.
In the case of diffractive multiparticle production we have to consider
rescattering effects in pomeron-hadron and pomeron-pomeron interactions
of enhanced graphs.
Whereas it was sufficient to introduce 
a renormalized pomeron trajectory to calculate cross sections, 
one needs for the calculation of particle production a model for 
the physical final
states which correspond to the unitarity cut of such a renormalized
pomeron propagator.
Following Refs.~\cite{Cardy74a,Kaidalov86d} we assume that  the 
pomeron-pomeron coupling can be described by the formation of an intermediate
hadronic system $h^\star$ where the pomerons couple to. Assuming
furthermore that
this intermediate hadronic system has properties similar to a pion,
the $n$-$m$ pomeron coupling  reads \cite{Kaidalov86d}
\begin{equation}
g_{n-m} = G \prod_{i=1}^{n+m-2} g_{h^\star\Pom}
\label{n-m-coupling}
\end{equation}
with $g_{h^\star\Pom} = g_{\pi\Pom}$ being the pomeron-pion coupling.
 $G$ is a scheme-dependent constant.
Hence, pomeron-hadron and pomeron-pomeron scattering 
should exhibit features similar 
to pion-hadron and pion-pion scattering.



To introduce hard interactions in diffraction dissociation, the exchanged
(renormalized) 
pomerons in pomeron--hadron and pomeron--pomeron scattering are again
treated as two-component objects
\begin{equation}
a_{A\Pom}(s,\vec B) \approx 
\frac{i}{2}\ G\ \left\{ 1 - \exp\left[-\chi_{\rm S}^{\rm diff}
-\chi_{\rm H}^{\rm diff} \right] \right\}
\label{two-comp-diff0}
\end{equation}
with the diffractive eikonal functions
\begin{eqnarray}
\chi_{\rm S}^{\rm diff} &=& 
\frac{g_{A\Pom}^0
g_{h^\star\Pom}^0 (M_D^2/s_0)^{\Delta_\Pom}}{8 \pi
b_\Pom(M_D^2)}
\nonumber\\
&\times&
\exp\left( -\frac{\vec{B}^2}{4 b_\Pom(M_D^2)}\right) 
\label{two-comp-diff1}
\\
\chi_{\rm H}^{\rm diff} &=&
\frac{\sigma_{\rm hard}^{A\Pom}(M_D^2)}{8 \pi
b_{\rm h,diff}}\exp\left( -\frac{\vec{B}^2}{4 b_{\rm h,diff}}\right).
\label{two-comp-diff2}
\end{eqnarray}
In all calculations the pomeron PDFs proposed by 
Capella, Kaidalov, Merino, and Tran (CKMT) \cite{Capella95a,Capella96a} 
with a hard gluon component are used.

\begin{figure}[thb] \centering
\epsfig{figure=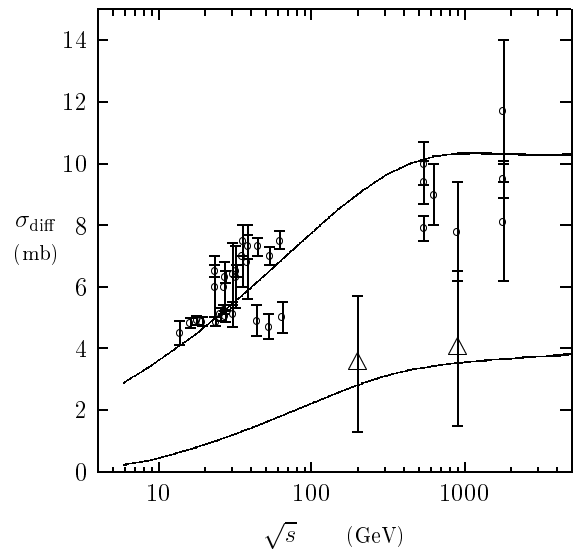,width=7.5cm}
\caption{
Single and double diffractive $p\bar p$ cross sections as a
function of the center of mass energy $\protect\sqrt s$ calculated with
the model.
We compare to data on single diffractive cross sections
\protect\cite{Chapman74,Schamberger75,Albrow76,Armitage82,%
Ansorge86,Robinson89,Amos90a,Amos93a,Abe94c}. 
In addition, some experimental
estimates for the cross section on double diffraction dissociation
\protect\cite{Ansorge86,Robinson89} are shown.
\label{ppdif}
}
\end{figure}

\begin{figure}[thb] \centering
\epsfig{figure=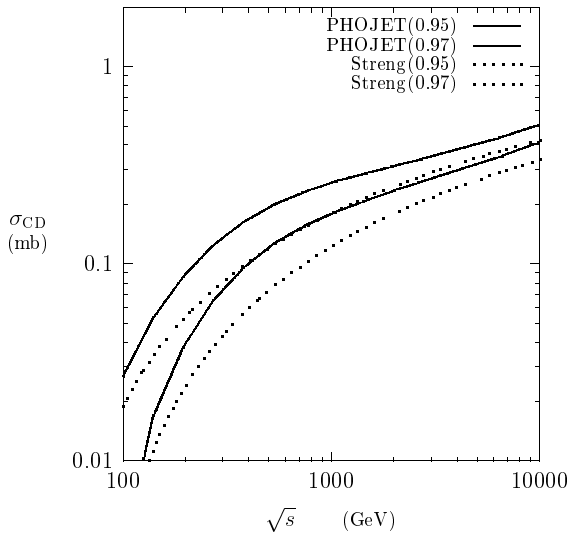,width=7.5cm}
\caption{
The energy dependence of the central diffraction
cross section. We compare the cross section as obtained from
\protect{\sc Phojet} with unitarization using a supercritical
pomeron with the cross section obtained by Streng
\protect\cite{Streng86a} without unitarization and with a critical
pomeron. Both cross sections are for the same two kinematic cuts:
$M_{\rm CD} > $2GeV/c${}^2$ and $c$ = 0.95 and 0.97.
The cross sections decrease with rising $c$.
\label{cdif}
}
\end{figure}


\subsection{Toy model with direct pomeron coupling}

To estimate the sensitivity of the model results to non-factorizing
coherent pomeron 
contributions as proposed in \cite{Collins93,Collins95a}, we 
use optionally also a toy model with a
direct pomeron-quark coupling \cite{Kniehl94}. 
In this case, the pomeron is treated similar
to a photon having a flavor independent quark coupling $\lambda$.
For definiteness, the corresponding matrix elements are given
\begin{eqnarray}
\left|M_{\Pom q \rightarrow\ q g}\right|^2 &= &
\lambda\alpha_s \left[
-\frac{8}{3}\frac{\hat{u}^2+\hat{s}^2}{\hat{s}\hat{u}}
\right]
\\
\left|M_{\Pom g \rightarrow\ q \bar q}\right|^2 &= &
\lambda\alpha_s\left[
\frac{\hat{u}^2+\hat{t}^2}{\hat{t}\hat{u}}
\right]
\\
\left|M^2_{\Pom \gamma \rightarrow\ q \bar q}\right|^2 &= &
\lambda \alpha_{\rm em} e_q^2\left[
6 \frac{\hat{u}^2+\hat{t}^2}{\hat{u}\hat{t}}
\right]
\\
\left|M_{\Pom \Pom \rightarrow\ q \bar q}\right|^2 &= &
\lambda^2\left[
6 \frac{\hat{u}^2+\hat{t}^2}{\hat{u}\hat{t}}
\right]
\end{eqnarray}
Here, $\alpha_s$ ($\alpha_{\rm em}$) denotes the strong
(electromagnetic)
coupling and $\hat s$,
$\hat t$ and $\hat u$ are the Mandelstam variables of the partonic
scattering process.

%


\section{Comparison with data\label{comparison-diff}}


\subsection{Diffractive cross sections}

First we compare single  diffractive
cross sections according to our model 
in $p$--$\bar p$ collisions to data and we present the results of
the model for single and double diffractive cross sections in
$\gamma$--$p$ collisions and for central diffraction cross sections
 in $p$--$p$ collisions.
Studying diffractive cross sections is not the primary concern
of this paper. Results on diffractive cross sections were
already presented  using the {\sc Dtujet} model in
Refs.~\cite{Aurenche92a,Bopp94a} and using the present {\sc
Phojet} model in Refs.~\cite{Engel95a,Engel96e}, we include
updated results for these cross sections here 
to make the present paper self-contained.

In Fig.~\ref{ppdif} data on single diffractive cross sections
\cite{Chapman74,Schamberger75,Albrow76,Armitage82,%
Ansorge86,Robinson89,Amos90a,Amos93a,Abe94c}
are compared with our model results. It is to be noted that the
data on single diffractive cross sections at collider energies
are subject to large uncertainties. Nevertheless the rise of the cross
section from ISR energies to the energies of the CERN and
FERMILAB colliders is less steep than expected from the Born level
expression from the triple pomeron formula (\ref{triple-Pom}).
It is the eikonal unitarization procedure in the model, which
suppresses the strong rise of the triple pomeron cross section in
the full model. The same effect was also found by Capella et al.\
\cite{Capella76} and Gotsman et al.\ \cite{Gotsman95a}.

In Fig.~\ref{cdif} we compare as function of the energy 
the central diffraction cross
sections in proton-proton collisions, 
which we obtain from {\sc Phojet} with the cross
section obtained by Streng \cite{Streng86a}. 
In {\sc Phojet} we use a supercritical pomeron with
$\Delta_{\Por}$ = 0.08 whereas Streng \cite{Streng86a}
uses a critical Pomeron with $\Delta_{\Pom}$ = 0.
Note that also the double-pomeron cross section 
grows in Born approximation
with $s$ like 
$\sim s^{2\Delta_\Por}$. This rapid increase is damped 
in {\sc Phojet} by the unitarization procedure. At high energies, 
contributions from multiple interactions become important. 
The  rapidity gaps are filled with hadrons due to 
inelastic rescattering and the cross section for central diffraction
gets strongly reduced. In contrast, Streng
calculates only the Born term cross section.
Figure~\ref{cdif}
illustrates the  differences obtained using 
different theoretical methods.
We stress, both methods use the measured
single diffractive cross sections to extract the triple-pomeron
coupling.


\subsection{Single diffraction in hadron-hadron collisions at
collider energies}

\begin{figure}[thb] \centering
\epsfig{figure=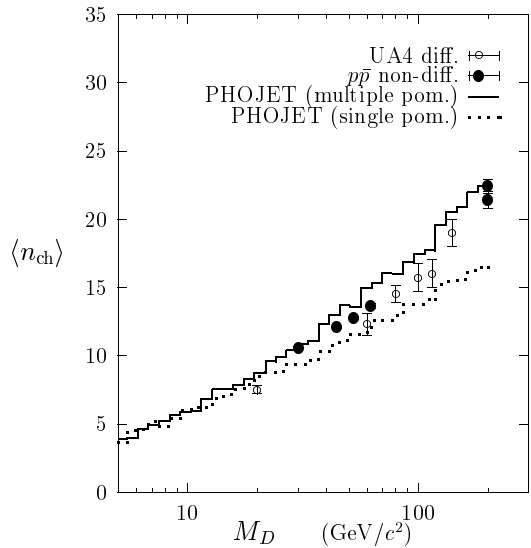,width=7cm}
\caption{
\label{ua4mul}
Mean charged particle multiplicity of the diffractively produced
hadronic system with invariant mass $M$. UA--4 data
\protect\cite{Bernard86a} are compared to single and
multiple interaction model predictions and data on non-diffractive
$p\bar p$ interactions at $\protect\sqrt{s}=M$.
}
\end{figure}
\begin{figure}[thb] \centering
\epsfig{figure=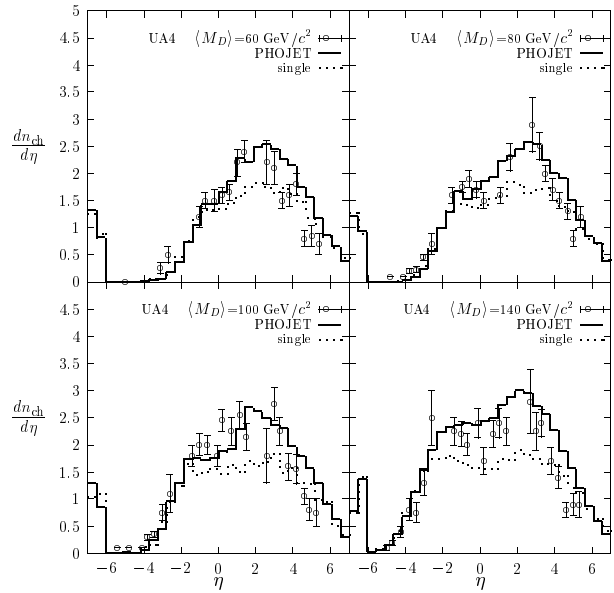,width=8cm}
\caption{
\label{ua4eta-c}
Pseudorapidity distribution of charged hadrons in single diffraction
dissociation. UA--4 data \protect\cite{Bernard86a} to
model predictions.
}
\end{figure}
There are the following experiments which have studied hadron production
in single diffraction
in $p\bar p$ collisions at the CERN--SPS--Collider:
\begin{enumerate}
\item
The UA--4 Collaboration \cite{Bozzo84b,Bernard86a,Bernard87b}
measured pseudorapidity distributions of charged hadron
production for different masses of the diffractive system. We
have already twice compared earlier versions of the 
Dual Parton Model\cite{Ranft87c,Roesler93} 
to this data.  New in the present model is 
hard diffraction and multiple chains in the diffractive hadron
production, 
therefore we have again compared to this data
and we find reasonable agreement (see Figs.~\ref{ua4mul} and
\ref{ua4eta-c}).
In particular we present besides the distributions
according to the full model also the contribution from one pair
of chains only (single interaction model). 
This is the rapidity distribution expected from
the Born term  without the contributions
from hard diffraction (minijets) and multiple soft interactions,
which are obtained from the unitarization method.
It is evident
from the data as well as from the model that multiple
interactions and minijets lead to a rising rapidity plateau in
pomeron--proton collisions in a similar way as observed 
in hadron--hadron collisions.
\item
Hard diffractive proton--antiproton interactions were
investigated by the UA--8 Collaboration \cite{Brandt92}. In this
experiment the existence of a hard component of diffraction was
demonstrated for the first time. Because of the importance of
these findings, we compared them already in a recent paper
\cite{Engel95c} to
our model and found the model  to be consistent with this experiment. 
Therefore we will not repeat this comparison here.
\end{enumerate}
%
\begin{figure}[thb] \centering
\epsfig{figure=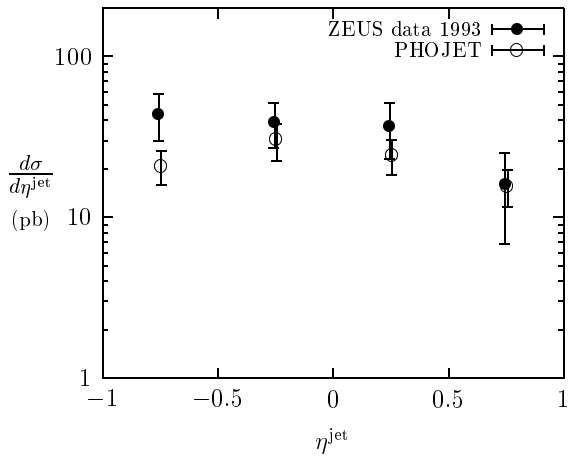,width=7.5cm}
\caption{
\label{zeus1}
Differential $e-p$ cross section $d\sigma /d\eta_{\rm jet}
(\eta^{had}_{max} < 1.8)$ for inclusive jet production with
$E_T^{\rm jet} >$ 8 GeV in the kinematic region $Q^2 \leq $ 4
GeV${}^2$ and 0.2 $< y <$ 0.85. We compare data from the ZEUS
Collaboration \protect\cite{Derrick95h} with \protect{\sc Phojet} results
using the same trigger as used for the ZEUS data.
}
\end{figure}
%
%
\begin{figure}[thb] \centering
\epsfig{figure=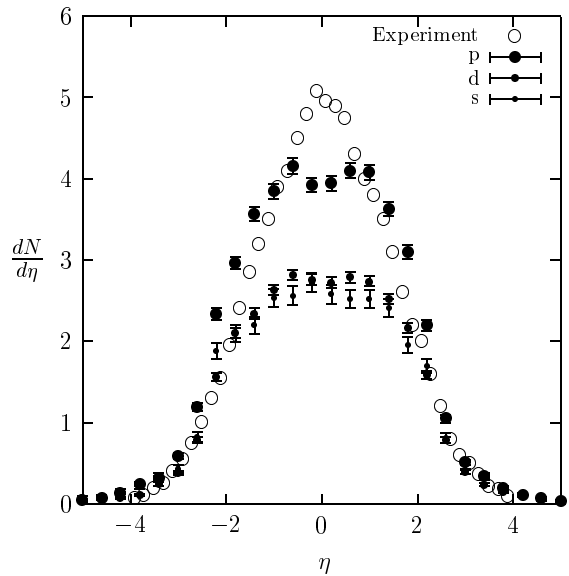,width=7.5cm}
\caption{
\label{ua1}
The pseudorapidity  distribution in central 
diffraction as observed by the
UA--1 Collaboration \protect\cite{Joyce93a} compared with the corresponding
distribution in \protect{\sc Phojet} without direct pomeron coupling
with  the UA--1 trigger applied to the Monte Carlo
events (p), with a direct pomeron coupling (d) and without
multiple interactions (s).
}
\end{figure}
%


\subsection{Single diffraction in photoproduction}

Results on single photon diffraction dissociation
and in particular hard single
diffraction were presented by both experiments at the HERA
electron--proton 
collider
\cite{Ahmed95a,Aid95b,Derrick95a,Derrick95h,Derrick95i,Derrick96b}.

The ZEUS Collaboration\cite{Derrick95h} 
has presented differential and integrated
jet pseudorapidity cross sections for jets with $E^{\rm jet}_T >$ 8
GeV. The absolute normalization of these data is given. This
allows one a more severe check of the model. In Figs.~\ref{zeus1} 
 we compare the differential  jet
pseudorapidity cross sections from ZEUS 
\cite{Derrick95h} 
to the model. The Monte Carlo events
 from {\sc Phojet}  have been treated with the
same cuts and trigger as used for the data. We find a reasonable
agreement. We should, however, point out that the data include
contributions from  non-diffractive processes while the results
from the model concern only diffractive events.


\subsection{Central diffraction dissociation}

Data on hard central  diffraction in proton--antiproton
collisions at 0.63 TeV have been published by
Joyce et al. \cite{Joyce93a}. These data were obtained with the
UA--1 detector at the CERN--SPS collider. The data are not easy
to understand since they have been obtained with triggers
demanding a pair of jets with $E_t > $ 3 GeV or localized
electromagnetic energy depositions larger than 1.2 GeV. This
trigger accepts a cross section of 0.3 $\mu$b while we find in
our model at this energy a total central  diffraction
cross section of approximately 0.3 mb (see Fig.~\ref{cdif}).
Thus the  trigger of Joyce et al.\cite{Joyce93a}
accepts only a tiny fraction of all central  diffraction
events. 
The most remarkable features of the data are the following:

\noindent
The
pseudorapidity distribution of the events accepted by the
trigger reaches a maximum central plateau of around 5 per
pseudorapidity unit, 30 percent higher than the non-diffractive
minimum bias events at the full $p$--$\bar p$ collision energy.

We try to understand these  data
\cite{Joyce93a} in three versions of the model.
(i) The full model without a direct pomeron
coupling, 
(ii) the full model with a direct pomeron quark coupling,
(iii) the model without multiple interactions and without a
direct pomeron coupling.
We use for the Monte Carlo events the same trigger requirements
as described in \cite{Joyce93a}. 

In Fig.~\ref{ua1} 
the charged particle $\eta$ distribution of the three versions of
the model are compared to the data. Only the full model gives a
pseudorapidity maximum comparable to the data. This is easy to
understand, only in the full model we have enough multiple soft
chains and multiple minijets to obtain such a large particle
density. In the model with direct coupling we trigger to events
with one pair of direct jets, this does not give enough particle
density. Similarly in the model without multiple interactions we
just get one pair of soft chains together with a minijet, also
in this configuration the particle density is lower than in the
full model.

\begin{figure}[thb] \centering
\epsfig{figure=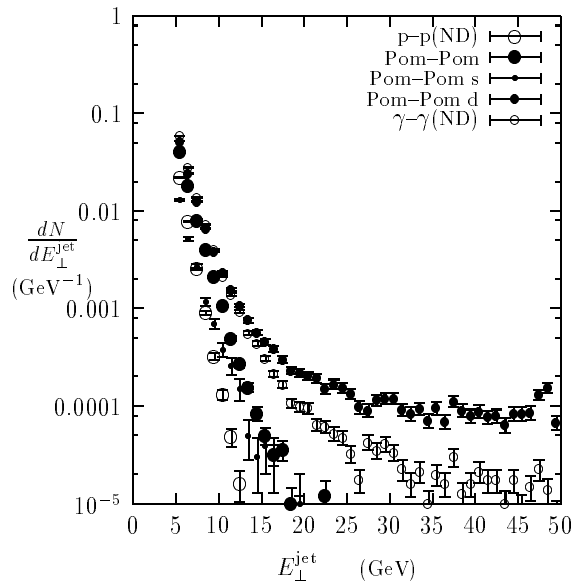,width=7.5cm}
\caption{
Jet transverse energy  distributions in non-diffractive $p$--$p$ and
$\gamma$--$\gamma$ 
 collisions compared with the jet transverse
energy  distribution in 
central  diffraction (pomeron--pomeron 
 collisions). For the latter channel we give the distributions
 separately for the full model, the model without multiple
 interactions (s) and the model with a direct pomeron coupling
 (d). The distributions
were generated with \protect{\sc Phojet}, the c.m.\ energy / diffractive
mass is 100 GeV in all cases.
\label{pt100jpopo}
}
\end{figure}
%
%
\begin{figure}[thb] \centering
\epsfig{figure=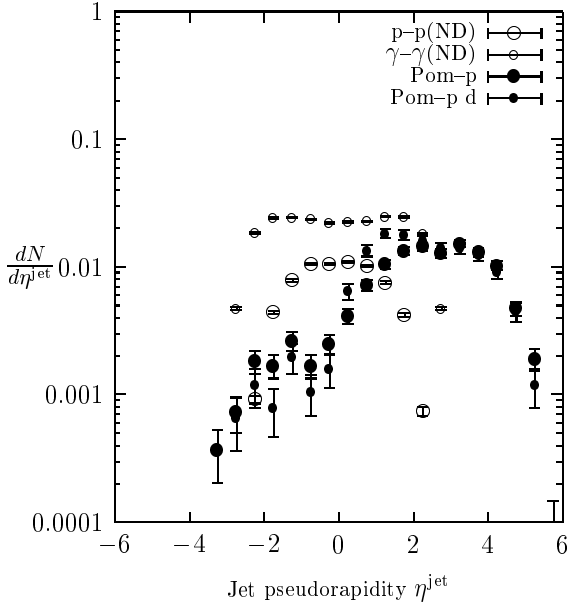,width=7.5cm}
\caption{
Jet pseudorapidity  distributions in non-diffractive $p$--$p$ and
$\gamma$--$\gamma$ collisions compared with the jet pseudorapidity
  distribution in 
 single diffraction (pomeron--$p$ 
scattering). The distributions
were generated with \protect{\sc Phojet} , the c.m.\ energy is 100 GeV in
all  cases, but the pseudorapidities  in the collisions with
pomerons given refer to the $\protect\sqrt
s$ = 2 
TeV $p$--$p$  collisions used to generate the
diffractive events.
\label{et100jpopo}
}
\end{figure}
%
%
%
\begin{figure}[thb] \centering
\epsfig{figure=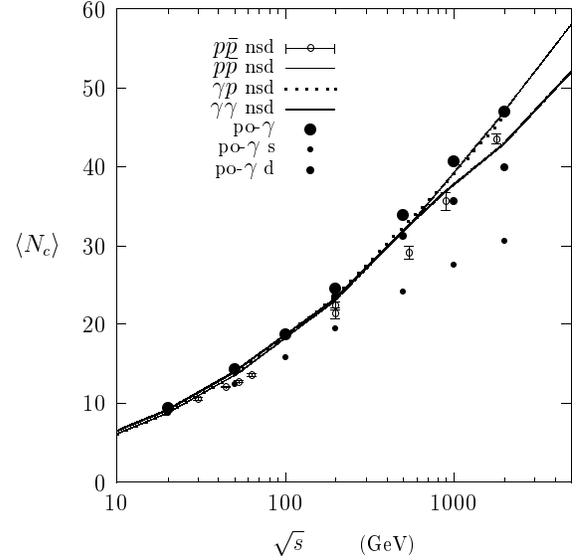,width=7.5cm}
\caption{
Average charged multiplicity as function of the c.m.\ energy
in single diffractive collisions
(pomeron--$\gamma$  collisions) according to
\protect{\sc Phojet} (points) is compared to the average charged
multiplicities in non single diffractive $p\bar p$, $\gamma$p and
$\gamma \gamma$ collisions, also according to \protect{\sc Phojet} 
(lines) and
experimental data in $p\bar p$ collisions.
\label{chmulpog1}
}
\end{figure}
%
%

\section{Comparing hadron production in diffractive processes to
non-diffractive particle production in $p$--$p$  and $\gamma$--$\gamma$
reactions\label{comparison-channel}}

In   Sections II  we have already pointed out, that our model for
particle production in pomeron--hadron/photon collisions and
pomeron--pomeron collisions has the same structure characterized
by multiple soft collisions and multiple minijets like models
for hadron production in hadron--hadron collisions. Therefore,
again we expect the main differences in comparison to other channels in
the hard component due to the differences between the pomeron
and hadron structure functions and
due to the existence or nonexistence of a direct
pomeron--quark coupling. We will use in all comparisons here
three models for $\Pom$--$p$, $\Pom$--$\gamma$ and $\Pom$--$\Pom$
collisions: 

\noindent
(i) our model with multiple soft and hard
collisions, 

\noindent
(ii) in order to see the influence of the multiple
soft and hard collisions a model with only one soft or hard
collision allowed and 

\noindent
(iii) the full model (i) assuming in addition
the existence of a direct pomeron--quark coupling according to
the toy--model . We present this despite
the fact that we did not find in the presently existing data any feature
which could only be described with such a coupling.


The differences
in the parton structure functions of protons, photons and
pomerons lead to quite different energy dependences of the hard
cross sections. In all processes where pomerons are involved,
single diffraction and central diffraction, hard
processes become important already at lower energies. For
pomeron--pomeron scattering at low energy the hard cross section
is about a factor 100 bigger than in $p$--$\bar p$ collisions. At
high energies the opposite happens, the hard cross sections in
all processes where pomerons are involved rise less steep with
the energy than in pure hadronic or photonic processes. The
reason for this is the different low-$x$ behavior of the
parametrization of the structure functions used. However, nothing
is known at present from experiment 
about the low-$x$ behavior of the pomeron
structure function.


In Fig.~\ref{pt100jpopo} we compare jet transverse
energy distributions  in $p$--$p$ and
$\gamma$--$\gamma$ collisions with the ones in 
$\Pom$--$\Pom$ collisions.  In the  channels with pomerons
we present again the distributions according to our full model, 
according to the model without multiple interactions and the
model with a direct pomeron--quark coupling. 
In  all non-diffractive collisions we have  $\sqrt s$ =
100 GeV and the diffractive events are generated in $\sqrt s $ = 2
TeV collisions with $M_D = 100$ GeV. 
The differences in the jet transverse energy
distributions between the channels
are as to be expected 
more important than in the hadron $p_{\perp}$ distributions.
We observe an important reduction in the jet distributions in
the model without multiple interactions. The effect of the
direct pomeron coupling is as dramatic as the effect due to the
direct photon coupling. The $E_{\perp}$ distributions in the
$\Pom$--$\gamma$ and $\Pom$--$\Pom$ channels extend up to the
kinematic boundary. In the latter two cases 
as in the case of $\gamma$--$\gamma$ 
collisions the entries at large $E_{\perp}$ come
only from direct processes.

In Fig.~\ref{et100jpopo}  we compare jet 
pseudorapidity distributions in $p$--$p$,
$\gamma$--$\gamma$ and $\Pom$--p, again, all collisions at $\sqrt s$ =
 100 GeV with the diffractive events generated in $\sqrt s $ = 2
TeV collisions.
For the jets we observe  substantial
differences in the shape of the pseudorapidity distributions.

In Figs.~\ref{chmulpog1}  we compare the average
charged multiplicity in non-diffractive $\bar p$--$p$, $\gamma$--$\gamma$
and $\gamma$--$p$ collisions according to the model as function of
$\sqrt s$ with the charged multiplicity in the pomeron--$\gamma$
diffractive
channel as function of the invariant mass of the diffractive
system. In the same plots we compare also to data in the case of
$\bar p$--$p$ collisions. 
We find at
collision energies below say 500 GeV only small differences
between the channels. However, at energies above 1 TeV the
model with only one pomeron exchange (one-pomeron cut) in diffraction
dissociation (labeled with s) predicts a smaller average multiplicity
than observed in hadron-hadron or photon-hadron scattering.




\section{Single diffraction and central diffraction at
TEVATRON}
\begin{figure}[thb] \centering
\epsfig{figure=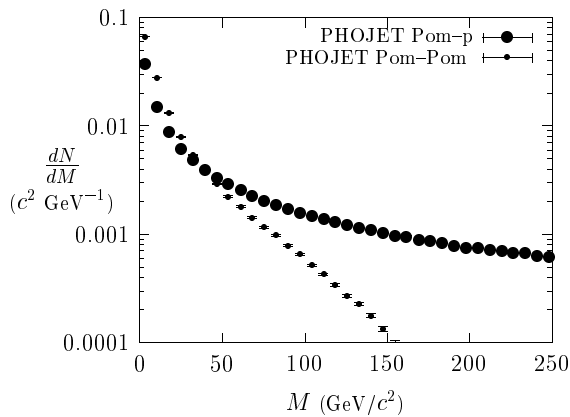,width=7.5cm}
\caption{
Distribution of the diffractive mass in single diffraction
(Pomeron--proton) and central diffraction (Pomeron--Pomeron) at
TEVATRON with $\sqrt s = 1.8$ TeV.
\label{fdndm}
}
\end{figure}

\begin{figure}[thb] \centering
\epsfig{figure=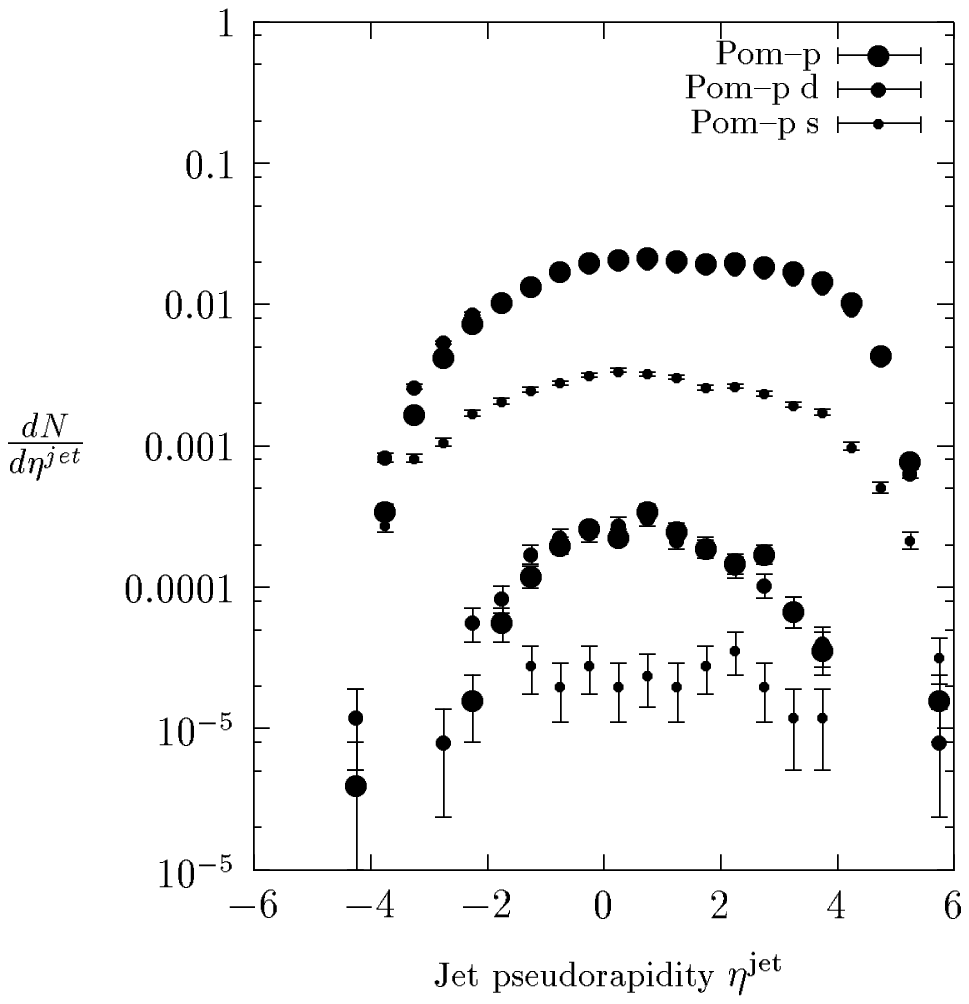,width=7.5cm}
\caption{
Pseudorapidity distribution of jets with $E_{\perp}$ larger
than 5 GeV  and 15 GeV 
in (one side) single diffraction (Pom--p) 
 at TEVATRON.
The upper curves with the same plotting symbol are generally 
for $E_{\perp}$ = 5 GeV, the lower curves are for $E_{\perp}$ = 15 GeV.
We plot also the distributions (d) using a small direct Pomeron coupling
($\lambda = 0.05$)  and (s) in a model where only single soft or
hard chains are permitted.
\label{fdndetaj}
}
\end{figure}

\begin{figure}[thb] \centering
\epsfig{figure=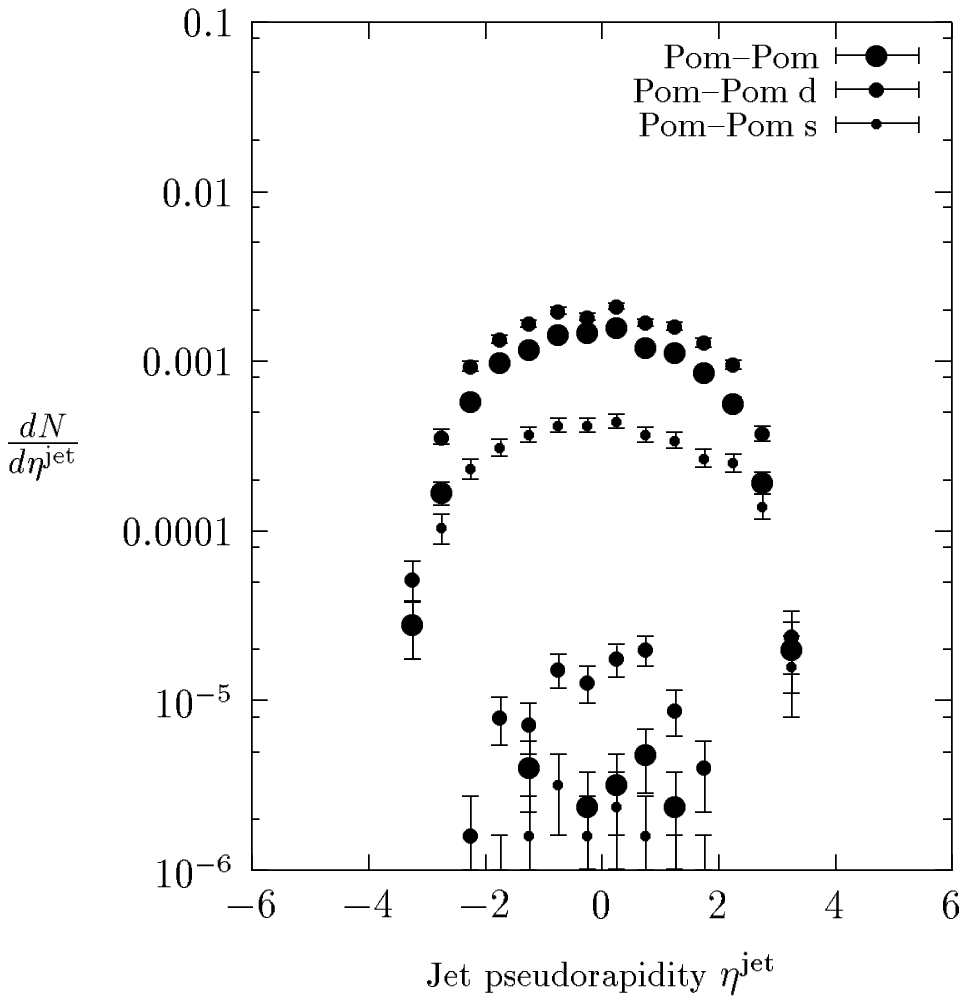,width=7.5cm}
\caption{
Pseudorapidity distribution of jets with $E_{\perp}$ larger
than 5 GeV  and 15 GeV 
in  central diffraction (Pom--Pom) at TEVATRON.
The upper curves with the same plotting symbol are
for $E_{\perp}$ = 5 GeV, the lower curves are for $E_{\perp}$ = 15 GeV.
We plot also the distributions (d) using a small direct Pomeron coupling
($\lambda = 0.05$)  and (s) in a model where only single soft or
hard chains are generated.
\label{fdndetajj}
}
\end{figure}

\begin{figure}[thb] \centering
\epsfig{figure=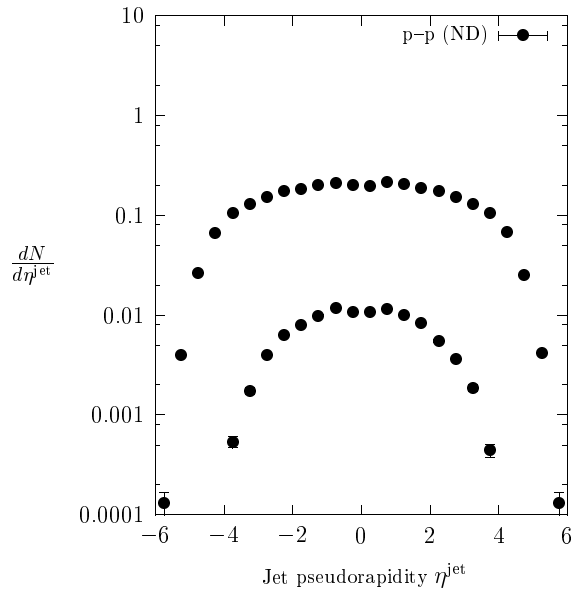,width=7.5cm}
\caption{
Pseudorapidity distribution of jets with $E_{\perp}$ larger
than 5 GeV  and 15 GeV 
in non-diffractive (ND) $p$--$p$ collisions 
 at TEVATRON.
The upper curve is
for $E_{\perp}$ = 5 GeV and the lower curve is for $E_{\perp}$ = 15 GeV.
\label{fdndetajjj}
}
\end{figure}

\begin{figure}[thb] \centering
\epsfig{figure=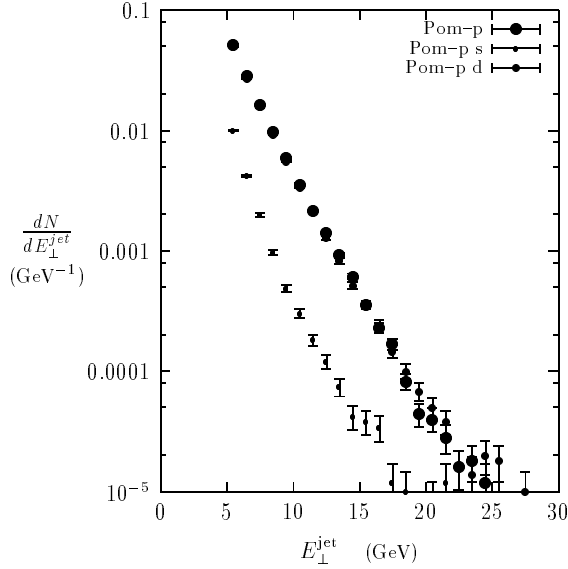,width=7.5cm}
\caption{
Transverse energy distribution of jets 
 in (one side) single diffraction (Pom--p) 
at TEVATRON.
We plot also the distributions (d) using a small direct Pomeron coupling
($\lambda = 0.05$)  and (s) in a model where only single soft or
hard chains are generated.
\label{fdndptj}
}
\end{figure}

\begin{figure}[thb] \centering
\epsfig{figure=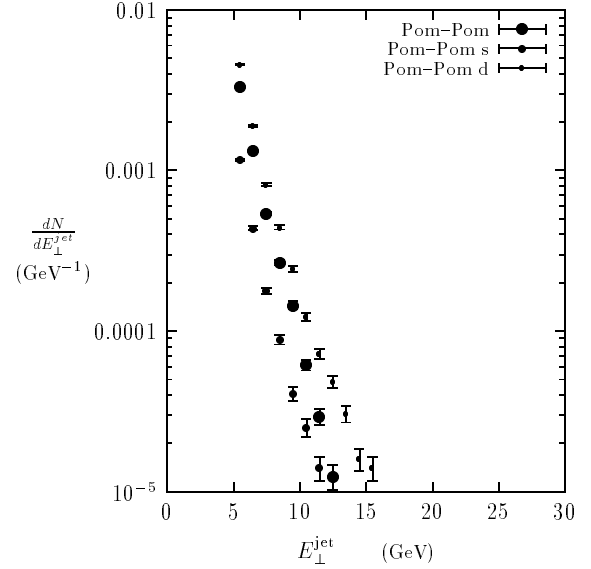,width=7.5cm}
\caption{
Transverse energy distribution of jets 
 in central diffraction (Pom--Pom) 
at TEVATRON.
We plot also the distributions (d) using a small direct Pomeron coupling
($\lambda = 0.05$)  and (s) in a model where only single soft or
hard chains are permitted.
\label{fdndptjj}
}
\end{figure}

\begin{figure}[thb] \centering
\epsfig{figure=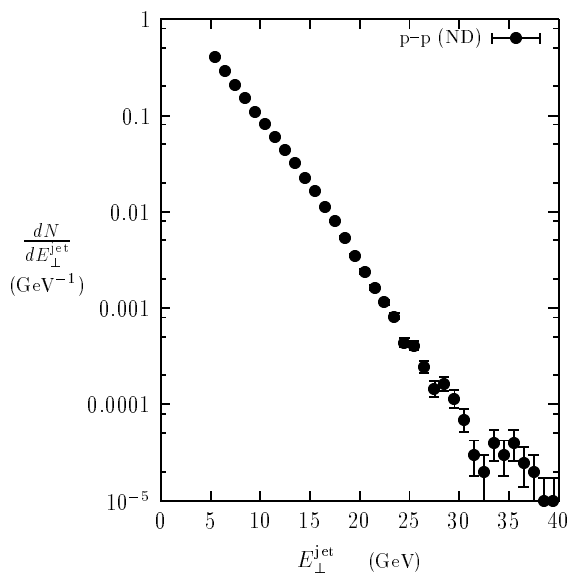,width=7.5cm}
\caption{
Transverse energy distribution of jets 
 in non-diffractive (ND) $p$--$p$ collisions at TEVATRON.
\label{fdndptjjj}
}
\end{figure}

\begin{figure}[thb] \centering
\epsfig{figure=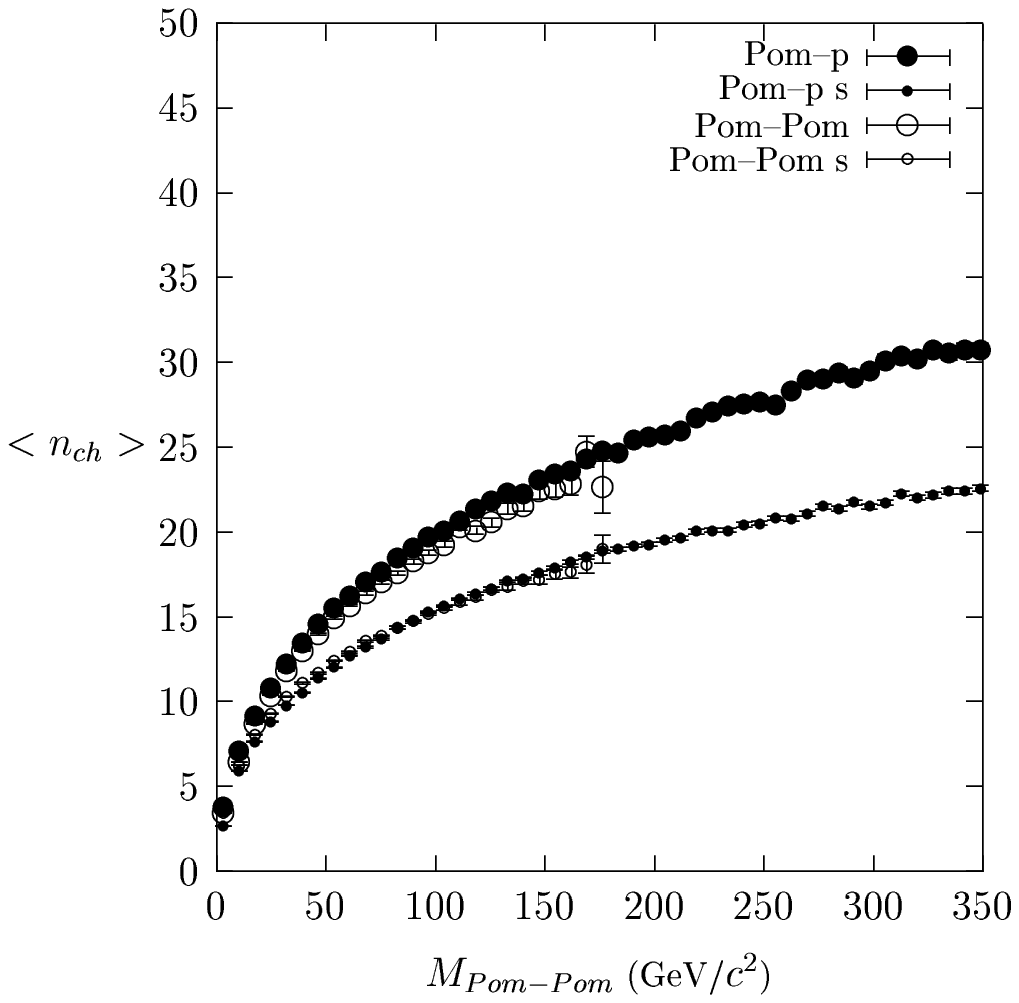,width=7.5cm}
\caption{
Charged multiplicity as function of the diffractive mass
 in single diffraction (Pom--p) and central
diffraction (Pom--Pom) at TEVATRON.
We plot also the distributions 
 (s) in a model where only single soft or
hard chains are considered.
\label{fnchm}
}
\end{figure}

In Figs. \ref{fdndm} to \ref{fnchm} we present some cross
sections calculated using {\sc Phojet} at TEVATRON energy. The
distributions are mass distributions in single and central
diffraction Fig. \ref{fdndm}, jet pseudorapidity distributions
in single and central diffraction  as well as in non-diffractive
$p$--$p$ collisions (ND) using  $E_{\perp}$
thresholds of 5 and 15 GeV Fig.\ref{fdndetaj} to
\ref{fdndetajjj}, Jet $E_{\perp}$
distributions Fig.\ref{fdndptj} to \ref{fdndptjjj} 
and the charged multiplicity as
function of the diffractive mass Fig.\ref{fnchm}. In some of
the distributions we give besides the full {\sc Phojet} model also the
plots for a model with a small direct pomeron coupling and for a
model with only single soft or hard chains pairs. 

Results on diffractive jet production from the two TEVATRON
Collaborations are discussed 
in \cite{Albrow97,Brandt97,Santoro97,Goul97,Brandt96}, 
one of the results
obtained by the D0 Collaboration is the ratio of 
double--pomeron exchange (DPE) (in the present paper we use
the term {\it central diffraction} (CD) instead of DPE) 
to non--diffractive (ND) dijet events:
\begin{equation}
\left(\frac{\sigma(DPE)}{\sigma(ND)}\right)_{E_{\perp}^{\rm jet}>15 GeV} 
\approx 10^{-6}
\end{equation}
{\sc Phojet} gives the following cross sections:

Non-diffractive (ND):$\sigma (ND) = $ 45.2 mb,

Single diffractive (SD):$\sigma (SD) =$ 11.2 mb,

Central diffraction (CD): $\sigma (CD) =$ 0.64 mb.

From these cross sections together with Figs.~\ref{fdndetaj} to 
\ref{fdndptjjj}  we get for this and similar ratios always for
$E_{\perp}$ larger than 15 GeV:

(CD)/(ND)$\approx 2 \times 10^{-6}$,

(SD)/(ND)$\approx 4 \times 10^{-3}$,

(CD)/(SD)$\approx 0.5 \times 10^{-3}$.

Despite the fact that no experimental acceptance has been considered for
these {\sc Phojet} results it is interesting to find the (CD)/(ND)
ratio so close to the D0 value given above.
%
%
%
%
%
%
%
%
%
%
%
%
%
%
%
%

\section{Conclusions and summary\label{summary}}

Multiple soft and multiple hard interactions (minijets)
have been introduced to describe high-mass diffractive hadron
production.
Comparing diffraction dissociation with the invariant mass $M$ to
non-diffractive particle production at $M=\sqrt{s}$,
a rise of the rapidity plateau and multiplicity is found
which is similar for both hadron production processes.
The model predictions agree well with data on high-mass single and central
diffraction dissociation.

Minimum bias hadron production in hadron-hadron,
and photon-photon collisions  as well as in pomeron--hadron,
pomeron--photon and pomeron--pomeron collisions of the same c.m.\
energy is remarkably similar. To see this, one has to restrict
the comparison to inelastic events and to exclude also the
diffractively produced vector mesons in reactions involving
photons. The only striking differences appear in the transverse
momentum distribution or distributions where the transverse
momentum behavior is essential. This difference can  be
understood to be due to the direct photon interaction
contribution and due to the photon and pomeron structure functions 
being considerably harder than hadronic structure functions. 

Finally we would
like to emphasize that  measurements at TEVATRON on CD and SD would
allow one to study  many of the open questions: Is it possible at
all to describe diffraction and hard diffraction 
using the triple pomeron graph? Can QCD factorization be applied to
the description of hard diffraction? Does a direct pomeron--quark
coupling exist? Do we have multiple soft and hard chains in
diffractive particle production?

\noindent
{\bf Acknowledgments}\\
The authors are grateful to F.W.~Bopp and S.~Roesler for many
discussions. One author (R.E.) thanks T.K.~Gaisser for helpful comments
and remarks.
The work of R.E.\ is supported in part by the
U.S.\ Department of Energy under Grant DE-FG02-91ER40626.


\bibliographystyle{zpc}

\bibliography{hep14}


\end{document}